\documentclass[useAMS,usenatbib]{mn2e}

\usepackage{graphicx}
\usepackage{amssymb}
\usepackage{pstricks}
\usepackage{longtable,lscape}
\usepackage{soul}
\def \be {\begin{equation}}
\def \ee {\end{equation}}

\def \be {\begin{equation}}

\def \ee {\end{equation}}

\title{Outwards migration for planets in stellar irradiated 3D discs}
\author[ E. Lega, A. Morbidelli, B. Bitsch, A. Crida, J. Szul\'{a}gyi ] { 
E. Lega$^{1}$, \thanks{E-mail: elena@oca.eu (EL);  morby@oca.eu (AM); bert@astro.lu.se (BB);  crida@oca.eu (AC);  jszulagyi@oca.eu (JS)},  A.Morbidelli ${^1}$,  B. Bitsch $^{2}$, A. Crida ${^1}$, J. Szul\'{a}gyi  ${^1}$ \\
$^{1}$ Universit\'e Nice Sophia Antipolis, CNRS UMR 7293,  Observatoire de la
C\^ote d'Azur,   Bv. de l'Observatoire, \\CS~34229,  06304 Nice cedex 4, 
France.\\
$^{2}$ Lund Observatory, Department of Astronomy and Theoretical Physics, Lund University, 22100 Lund, Sweden. }

\begin{document}
\maketitle
\date{Accepted .... Received ....; in original form ....}

\pagerange{\pageref{firstpage}--\pageref{lastpage}} \pubyear{2015}
\label{firstpage}

\begin{abstract}
  For the very first time we present 3D simulations of planets embedded in stellar irradiated discs. It is well known that  thermal effects could reverse the direction of planetary migration from inwards to outwards, potentially saving planets in the inner, optically thick parts of the protoplanetary disc. When considering stellar irradiation in addition to viscous friction as a source of heating,
the outer disc changes from a shadowed to a flared structure. Using a suited analytical formula it has been shown that in the flared part of the disc the migration is inwards;  planets can migrate outwards only in  shadowed regions of the disc, { because the radial gradient of entropy is stronger there}. In order to confirm this result numerically,  we have computed  the total torque acting on planets held on fixed orbits embedded in stellar irradiated 3D discs  using the  hydrodynamical code FARGOCA. We find qualitatively good agreement between the total torque obtained with numerical simulations and the one predicted by the analytical formula. For large masses ($>20M_{\earth}$) we find  quantitative agreement, and we obtain outwards migration regions for planets up to $60\mathrm{M_{\earth}}$ in the early stages of accretional discs.  { We find nevertheless that the agreement with the analytic formula is quite fortuitous because the formula underestimates the size of the horseshoe region; this error is compensated by imperfect estimates of other terms, most likely  the cooling rate and the saturation}.
\end{abstract}

\section{Introduction}
In recent years it has been shown that low mass planets ($[5,20]\mathrm{M_{\earth}}$)
can migrate outwards in discs with non-isothermal effects  \citep{PaardeMelle06,BarMass08,KleyCri08,KBK09,MasCas09,MassetCasoli10,Paaretal10,Paaretal11,LCBM14}.
Precisely, the migration in the inner 
part of a  radiative disc can be directed outwards, 
while it remains directed  inwards in the outer disc \citep{BK11}. This establishes the existence of a critical radius where
migration vanishes, towards which  planetary cores  migrate from both the inner and the outer part of the disc. Therefore, the zero  migration location acts as  a planet trap { at which proto-planets can accumulate in resonances, collide and eventually form bigger objects.} \citep{Lyraetal10,Cossouetal14,ColeNels14}. 
 
\par  In all these works the heating is provided by viscous friction and the cooling by radiative diffusion (in 3D discs) or by a local cooling rate (in 2D discs). More recently, \citet{BCMK13},  have shown the importance of stellar irradiation on the disc structure and the consequences for planetary migration.
The main result is that when considering stellar irradiation the outer disc changes from a shadowed to a flared disc. Nonetheless, opacity transitions (e.g. the iceline) create bumps in the aspect ratio and hence shadowed regions.
In the flared part of the disc the migration 
turns out to be inwards and planets can migrate outwards only in  shadowed regions of the disc, { where the radial gradient of entropy is much steeper than in the flared part of the disc}. These results were obtained  in equilibrium discs with uniform viscosity { featuring a zero radial mass flux} \citep{BCMK13} and extended to the  case of  accretion discs with an alpha prescription for the viscosity \citep{ShakuraSun1973} in \citet{PaperII}.  This second case is of particular interest, since a given mass-flow ($\dot M$) rate corresponds to a specific disc age \citep{Hartmann98}  so that, investigating the disc structure for different values of $\dot{M}$, is equivalent to studying the disc structure as a function of the disc evolution \citep{PaperII,BJLM15}.  

\par 
In \citet{BCMK13,PaperII,BJLM15} the planet migrations maps have been obtained applying the torque formula of \citet{Paaretal11} using the disc properties obtained in the simulations.  The formula, although very complex, is based on the fact that the torque exerted by the protoplanetary disc { onto the planet} has two main contributions:  i) the so-called Lindblad torque due to the spiral arms launched by the planet  in the disc, which is not affected by the equation of state \footnote{Actually, it scales with $\gamma \Gamma_0$, with $\Gamma_0$ given in Eq.\ref{gamma0}.},  and ii) the co-orbital corotation torque caused by material librating in the horseshoe region.  
When considering radiative effects the corotation
torque contribution can be positive and possibly dominate over the negative
Lindblad torque, { leading to outwards migration}. The Paardekooper et al. formula was calibrated with a 2D hydrodynamical model for low mass planets ($5\mathrm{M_{\earth}}$). 

\par With a specific disc setting not accounting for stellar irradiation, \citet{KBK09} found positive total torque for planetary masses in the
range $[5,30]$ Earth masses. The influence of the disc's mass on the migration was studied in \citet{BK11} and its application to Earth sized planets was addressed by \citet{LCBM14} who found the contribution of a new torque not accounted for by the formula. 
This new negative torque is due to the formation of an asymmetric cold and dense finger of gas driven by circulation and libration streamlines. 
\par
The aim of this paper is to investigate numerically the total torque acting on planets kept on fixed orbits in 3D stellar irradiated discs and provide quantitative comparisons with the \citet{Paaretal11}  formula.  We use the explicit/implicit hydrodynamical code FARGOCA \citep{LCBM14} that includes a two-temperature solver {  for radiative transfer} in the flux-limited approximation.

\par
 In principle stellar heating only changes the
structure of the disc and does not act on the mechanism responsible for 
outwards migration directly. Nevertheless  { the analytical formula has been calibrated on 2D discs  so that a quantitative test on the validity of that formula for realistic  3D discs is needed}. 
Moreover, a constant mass flow $\dot M$ through the disc \citep{PaperII} might have an influence on the torque acting on planets and, to our knowledge, this case has never been tested before in 3D  discs with a non-isothermal EOS.

\par The paper is organized as follows: the physical modelling is presented in Section 2, in Section 3 we describe the migration maps obtained from the analytic formula applied to our 2D unperturbed discs \footnote{discs in thermal equilibrium in the $(r-z)$ plane}. In Sections  4 and 5 we provide results of 3D simulations done respectively on a constant viscosity stellar irradiated equilibrium disc and on an $\alpha$-viscosity  accretion disc. { In section 6 we go beyond the simple comparisons of the net torques predicted analytically and measured numerically, focusing on a key ingredient of the analytic formula: the size of the corotation zone, which governs the corotation torque and the torque saturation.} The conclusions are provided in Section 7. 

\section{Physical Modelling}
The protoplanetary disc is treated as a three dimensional non self-gravitating gas whose motion is described by the Navier-Stokes equations.
We use spherical coordinates $(r,\theta, \varphi)$ 
where $r$ is the radial distance from the star, i.e from the origin, $\theta$ the 
polar angle measured from the $z$-axis (the colatitude) and $\varphi$ is the azimuthal coordinate
starting from the $x$-axis. The midplane of the disc is at the equator
$\theta = {\pi \over 2}$.
We work in a  coordinate system which rotates with angular velocity:
$$\Omega_p = \sqrt {G(M_{\star}+m_p) \over {a_p}^3} \simeq \sqrt {GM_{\star} \over {a_p}^3} $$
where $M_{\star}$ is the mass of the central star, { $G$ the gravitational constant}
and $a_p$ is the semi-major axis of a planet of mass $m_p$, assumed to be on a circular orbit.
{ The gravitational influence of the planet on the disc 
 is modelled as in \cite{KBK09}
using a cubic-potential of the form:
\begin{equation}
\Phi _p = \left\lbrace \begin{array}{ll}
-{m_pG\over d} &  d > \epsilon \\
-{m_pG\over d}f({d\over \epsilon}) & d\leq \epsilon   
\end{array} \right.
\label{cubic}
\end{equation}
where  $d$ is the distance from the disc element to the planet and $\epsilon$ is the softening length.
Writing $x=d /\epsilon$ the function $f$ is given by: 
 $$f(x) = x^4-2x^3+2x \ . $$ \par
We have considered $\epsilon = 0.6 R_H$ in our simulations sets SED and  SAD}
{ and  $\epsilon = 0.5 R_H$ for simulation set RED, with
$R_H$  the Hill radius of the planet:
$$R_H= a_p\sqrt[3]{m_p \over {3M_*}}\ .$$ We will discuss in Section 6 the dependence of the total torque on the softening length. \par In the Paardekooper formula, there is also a softening length parameter $\beta$
and it has been fixed to $\beta=0.35H$ ($H$ being the disc's local scale height) from previous comparison between the formula and 3D simulations \citep{BK11}. }

The hydrodynamical equations solved in the code are described in \citet{LCBM14}, the two temperatures approach for the stellar irradiation was described in detail in \citet{BCMK13} and the opacity prescription in \citet{PaperII}. The flux received from the star at a radial distance $r$ is:
\begin{equation}
F_{\star} = {R_{\star}^2\sigma T_{\star}^4/r^2}
\end{equation}
were the radius of the star is set to $R_{\star}=1.5\mathrm{R_{\sun}}$ and the temperature 
$T_{\star}=4370\mathrm{K}$  (appropriate for a Solar-type protostar). 
We consider the disc settings  of \citep{BCMK13}  namely a disc extending from
$r_{min}\leq r \leq r_{max}$ with $r_{min}=1 \mathrm{AU}$ and $r_{max}=50\mathrm{AU}$.
In the vertical direction the disc extends from the midplane ($\theta \simeq 90^{\circ} $) to $20^{\circ}$ above the midplane, i.e. $\theta \simeq 70^{\circ} $).
The initial surface density profile is $\Sigma(r)=\Sigma _0 (r/a_J)^{-b}$ and $a_J=5.2\mathrm{AU}$.
\par
We consider in the following different sets of simulations. Precisely, we consider a stellar irradiated equilibrium disc (set SED, hereafter), with uniform viscosity (viscosity coefficient 
$\nu = 10^{-5}a_J^2\Omega_p$), $\Sigma_0= 4.88\times 10^{-4}$ in code units ($147 \mathrm{g/cm^{2}}$)  and  $b=0.5$.  We also consider  a stellar irradiated  accretion discs  (set SAD, hereafter)  with alpha viscosity  ($\alpha=0.0054$), constant $\dot M=4\cdot 10^{-8}\mathrm{M_\odot/yr}$  and initial value of  $b=15/14$ and
of $\Sigma_0=430 \mathrm{g/cm^2}$. The initial value of $b$ corresponds to the radial surface density profile in flared disc \footnote {with flaring index $2/7$}  with constant $\dot M$ at all orbital distances. 
{ A disc with $\dot M=4\cdot 10^{-8}\mathrm{M_\odot/yr}$  corresponds to the early evolution stages of accretion discs and is probably younger than 1 Myr \citep{Hartmann98}.}
 The disc structure of the SAD case is discussed in \citet{BJLM15}.\par
 For comparison, we will also consider the case of a radiative equilibrium disc with uniform viscosity ($\nu=10^{-5}a_J^2\Omega_p$ ) and $b=0.5$, were the only source of heating is viscous heating (set RED hereafter). { This setup of the} RED disc has been previously studied in \citet{KBK09, BK11} and in \citet{LCBM14} for planetary masses up to $30 \mathrm{M_{\earth}}$.

\par
\begin{figure}
\includegraphics[height=6.5truecm,width=6.5truecm]{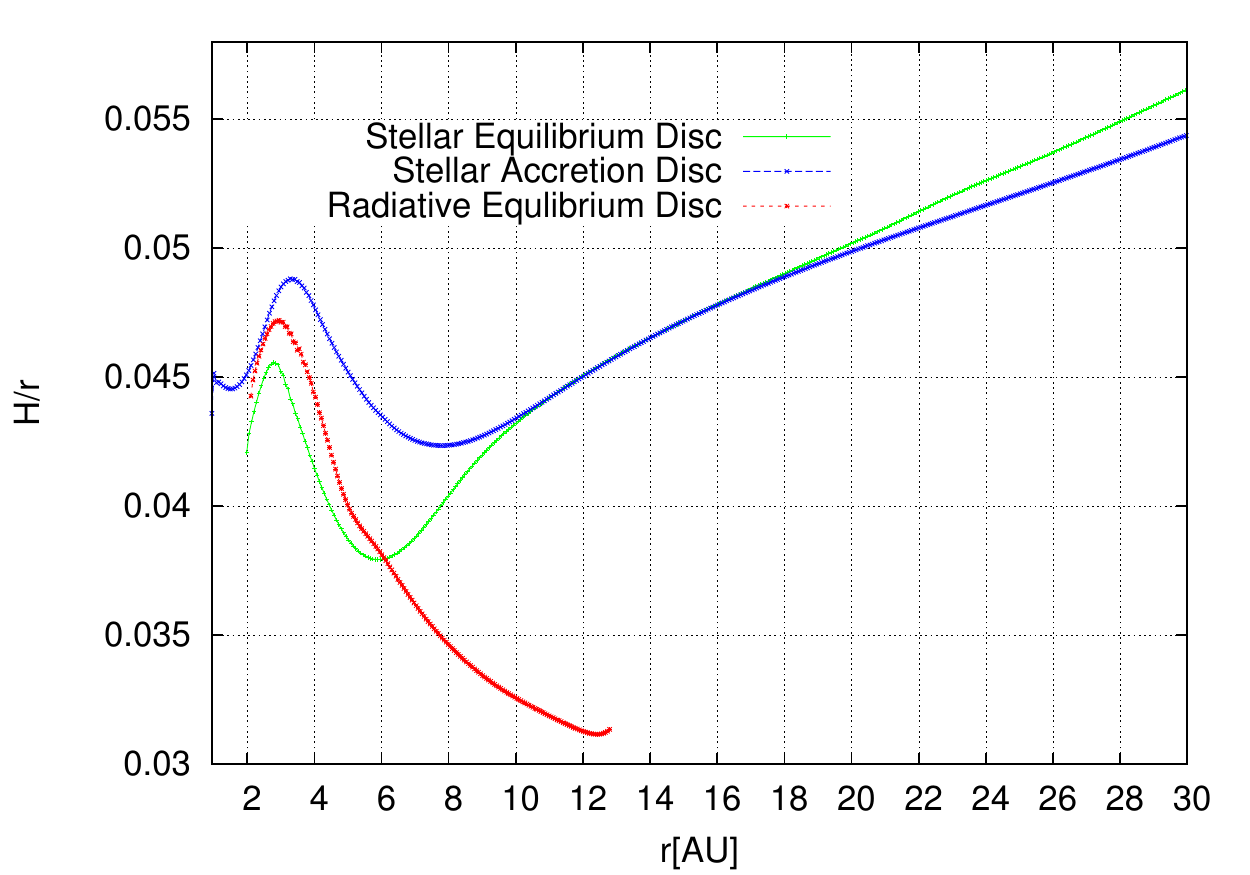}
\caption{Aspect ratio of the 2D ($r,\theta $) equilibrium 
 of simulation set SED (green) and SAD  with $\dot M=4\,10^{-8}$ (blue). For comparison the red curve provides
 the aspect ratio of the radiative equilibrium disc (RED) studied in
\citet{KBK09,LCBM14}.}
\label{hr}
\end{figure}

Before placing the planet in a 3D disc, we bring the disc to radiative equilibrium. 
We first model each disc in  2D, with  coordinates   $(r,\theta)$. For the
accretion disc the  $(r,\theta)$ disc evolution is explained in detail in \citet{PaperII,BJLM15}.
Once the 2D equilibrium is achieved all the gas fields are expanded to 3D.
The aspect ratio of  the 2D equilibrium for the SED, the SAD with $\dot M=4\,10^{-8}\mathrm{M_\odot/yr}$ and the RED  are reported in Fig.\ref{hr}. We remark that for the SED and RED  with uniform viscosity, the settling of the aspect ratio to an equilibrium value does not change the surface radial density  profile of the disc. Instead, in the SAD disc, in which the viscosity follows the alpha-prescription and thus depends on the disc's local scale height $H$, the resulting surface density profile at equilibrium is somewhat different from the initial one (see \citet{PaperII}). 

In the two cases where the disc is heated by the star we observe the typical profile of a flared disc, while we have a shadowed disc in the RED case.  The bump in $H/r$ around $3-4$ AU is caused by a transition in opacity that changes the local cooling rate of the disc and hence changes the disc's temperature \citep{PaperII}.
\par
The resolution of our computational grid is chosen in order to have  in the radial direction { approximatively $n$} grid-cells in the horseshoe region.
The half-width of the planet's horseshoe region is given, in the isothermal disc approximation \citep{MassetDangelo06}, by:
\begin{equation}
x_{hs2D}=1.16a_p\sqrt{q\over h} \ .
\label{xhs}
\end{equation}
were $q=m_p/M_{\star}$ and $h$ is the disc aspect ratio at $a_p$. 
We have checked that this formula { can still be used as a guideline for the choice of the resolution} for non isothermal 3D discs with and without stellar irradiation. { However, Eq.\ref{xhs} is based on the equivalence of the linear corotation torque and the horseshoe drag in 2D discs. In \citet{MassetDangelo06} it has been shown that nonlinearities appear on the flow for mass ratios 
$q\geq h^3$. For these planetary masses
  the horseshoe region is larger than the linear prediction  resulting in a boost of the corotation torque.  It is therefore
 important to check whether also in 3D hydrodynamical simulations the horseshoe
width increases for $q \geq h^3$ and what is the corresponding impact in the corotation torque. Section 6 is devoted to this topic. }
 \par
The values of the masses and  resolutions 
$(N_r,N_\theta,N_\varphi)$ for the SED disc 
are shown in Table \ref{table:tab1}. 
In \citet{LCBM14} we have shown that a resolution of 4 grid cells in the half width of the horseshoe region matched the requirement of having resolution independent results within reasonable CPU times for masses up to $20 \mathrm{M_{\earth}}$. Here we  use 4 grid cells in the half width of the horseshoe region also for planets with masses larger than $20 \mathrm{M_\oplus}$;  We have tested
that results are stable when increasing the resolution to 5  grid cells  in some test cases.

\par { Concerning the calculation of the gravitational torque acting on the planet  we recall that it is common to exclude the inner part of the Hill sphere of the planet.
This is obtained by applying a tapering function (named Hill cut in the following)
which in our case reads \citep{KBK09}:
\begin{equation}
f_b(d)=\left[ \exp \left(-{d/R_H -b \over b/10}\right)+1\right]^{-1}.
\label{Hcut}
\end{equation}
The value of $f_b$ is zero at the planet location and increases to 1 at distances  $d$ from the planet larger than the Hill radius $R_{H}$. The parameter $b$ 
denotes the distance from the planet, in unit of Hill radius, at which $f_b=1/2$. Here we use $b=0.8$ \citep{Crida09}.

\par
This procedure allows to exclude the part of the disc that is  gravitationally bound to planets that form a circum-planetary disc.
However, this prescription is not  justified
for small mass planets that do not have a circum-planetary disc.
{ For small planets we mean planets having a Bondi radius ($R_B$) smaller than their Hill radius. From the definition of the Bondi and Hill radius one obtains $R_B<R_H$ for $q < h^3/\sqrt 3$ \footnote{ We notice that the
parameter that determines the flow linearity in the planet vicinity is 
$q<h^3$ \citep{MassetDangelo06}.}.}
 In the following
we show the total torque computed with Hill cut, and,  for planetary masses satisfying {  $q < h^3/\sqrt 3$,} we also show the total torque computed without Hill cut.

\section{Migration maps}
The change in the disc structure due to stellar irradiation has important consequences for the migration of embedded bodies \citep{BCMK13}.
One can estimate the torque acting on planets  using the formula provided by \citet{Paaretal11}.
The migration maps obtained from the formula are  shown on Fig.\ref{map}
for respectively set SED (top panel), set SAD with $\dot M=4\,10^{-8}\mathrm{M_\odot/yr}$ (middle panel) and set RED  (bottom panel). The values of the total torque $\Gamma_{tot}$ are are  normalized with respect to
\begin{equation}
\Gamma_0= (q/h)^2\Sigma_p a^4_p\Omega^2_p
\label{gamma0}
\end{equation}
 where $\Sigma_p$ is the disc's surface density at the
planet location $a_p$. \par
{ We do not enter here in the details of the torque formula provided by 
\citet{Paaretal11}
we just recall that outwards  migration is a delicate process depending on the dynamical properties of  the corotation region as well as on the viscosity of the disc and of its cooling properties. 
Different timescales are at play (see  \citet{BK10}),
 in order to detect a positive corotation torque contribution which possibly dominates over the negative Lindblad torque. \par
 We remark that the formula has been obtained for planets that do not perturb the disc significantly ($q<h^{3}$, i.e. linear regime)  while the maps shown in Fig.\ref{map} consider also intermediate mass planets ($30-70 M_{\earth}$). The torque saturation, which determines the upper limit in planet mass for outwards migration, depends on the viscous timescale defined by:
 \begin{equation}
\tau_{\nu} = {(x_{hs2D}\gamma^{-1/4})^2/\nu}
\label{taunu}
\end{equation}
were $\gamma$ is the adiabatic index, $\gamma=1.4$ in our simulations. 
The width of the horseshoe region is  $x_{hs2D}/\gamma^{1/4}$ in order to take  adiabatic effects into account \citep{PaaPap09}.
However,  for intermediate mass planets, the actual width of the horseshoe region can be  different from $x_{hs2D}$ of Eq.\ref{xhs} \citep{MassetDangelo06} and this difference can have an impact on  torque saturation. \par
In the migration map of the SAD case (Fig.\ref{map},top) we see that, using
 $x_{hs2D}/\gamma^{1/4}$, torque saturation is supposed to occur for quite large planetary masses and  outwards migration is expected for planets up to $70M_{\earth}$ (this upper limit can moderately change
by changing the smoothing length $\beta$). \par
It is interesting to check this result with 3D numerical simulations, because outward migration of intermediate mass planets has important impacts for formation models of giant planets. In fact giant planet precursors could be prevented from migrating into the inner disk before they reach a mass that allows gap opening and therefore slower inward migration in the type-II migration regime.
}\par

In order to test the validity of the formula we have considered planets of different masses
held on fixed orbits in each of the three considered discs and we have computed the evolution of the torque with time
until we obtain  a stationary state.
We choose the distance of the  planets from the star  within the region of expected outwards migration from  Fig.\ref{map}. 
\begin{figure}

\includegraphics[height=6.5truecm,width=7truecm]{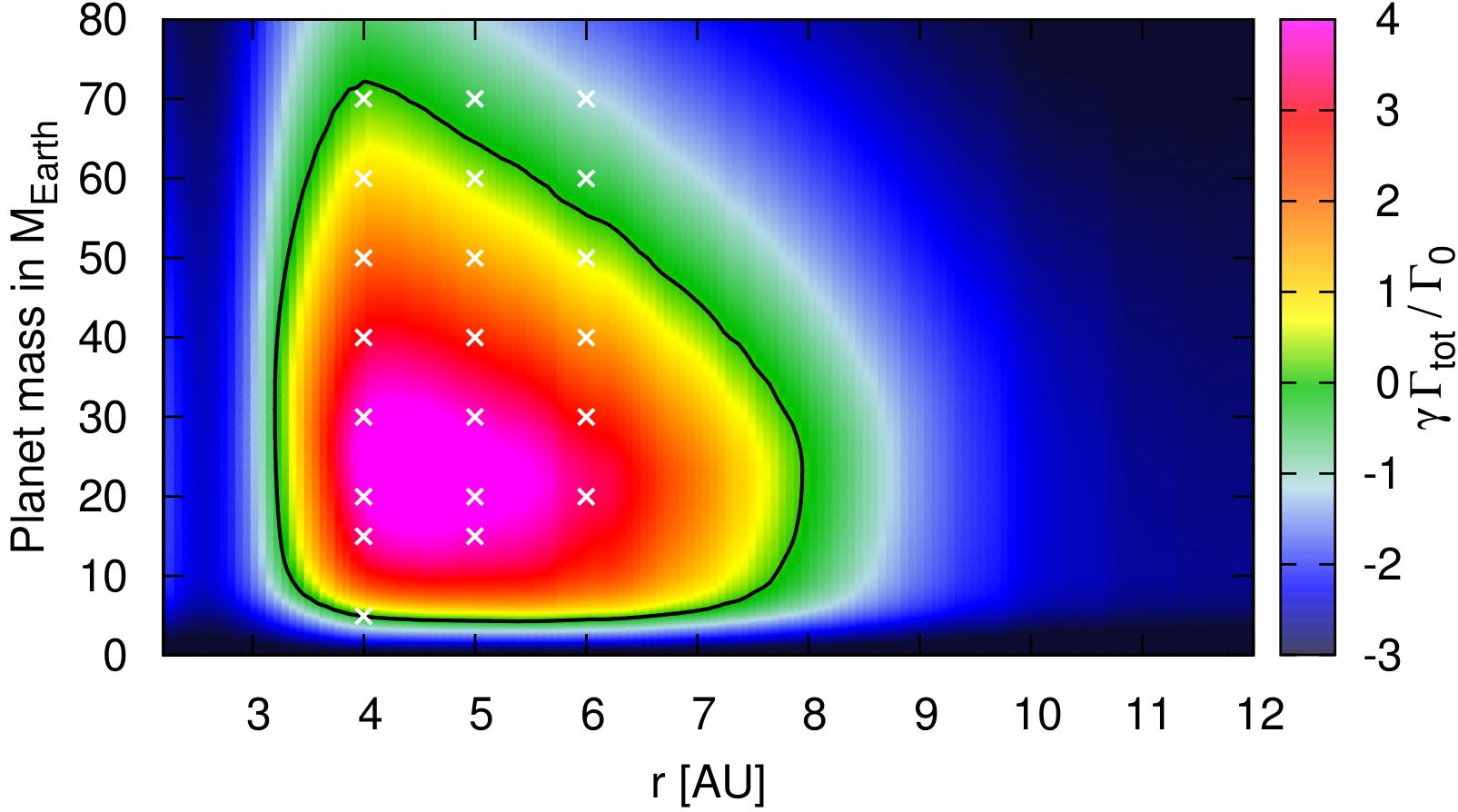}
\includegraphics[height=6.5truecm,width=7truecm]{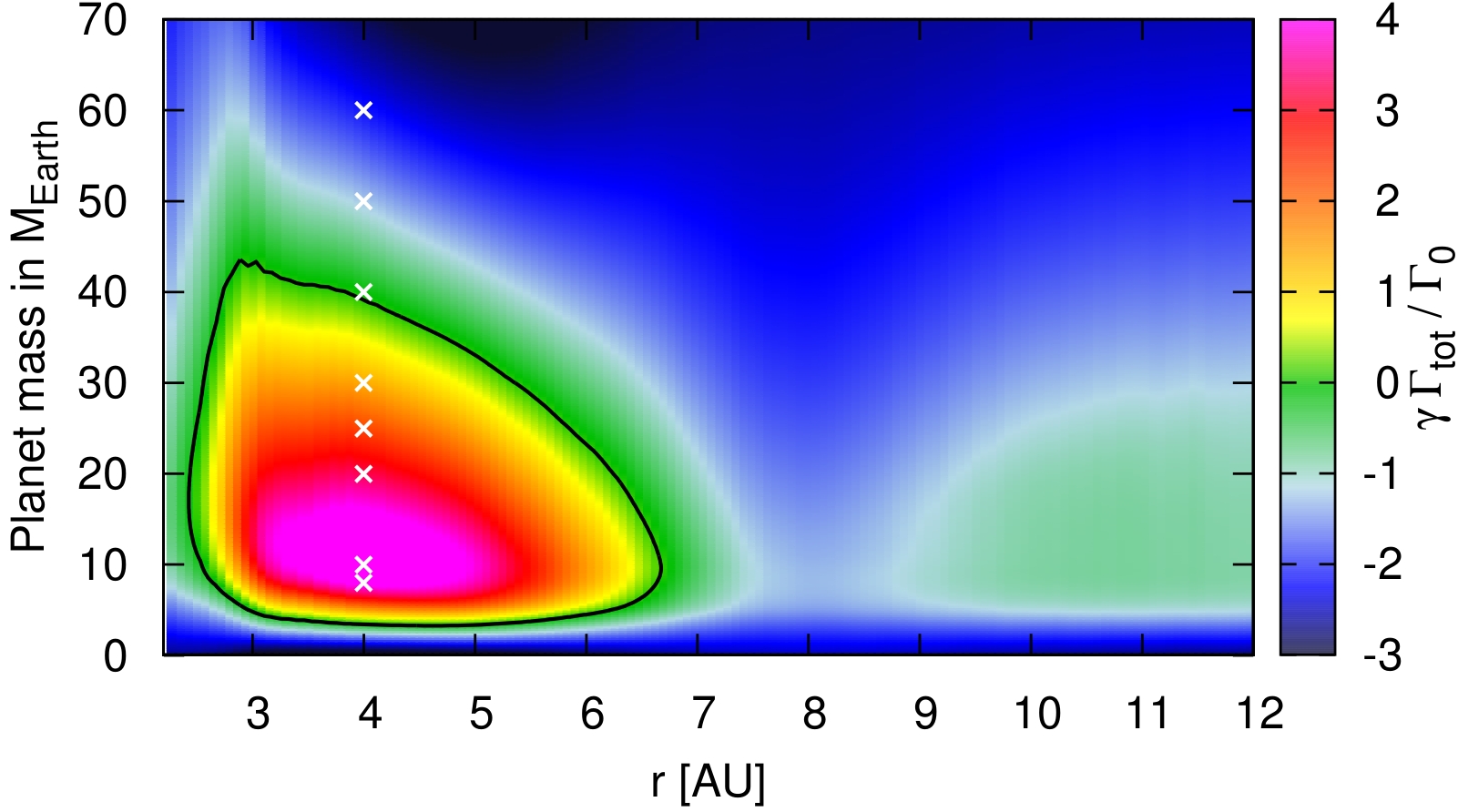}
\includegraphics[height=6.5truecm,width=7truecm]{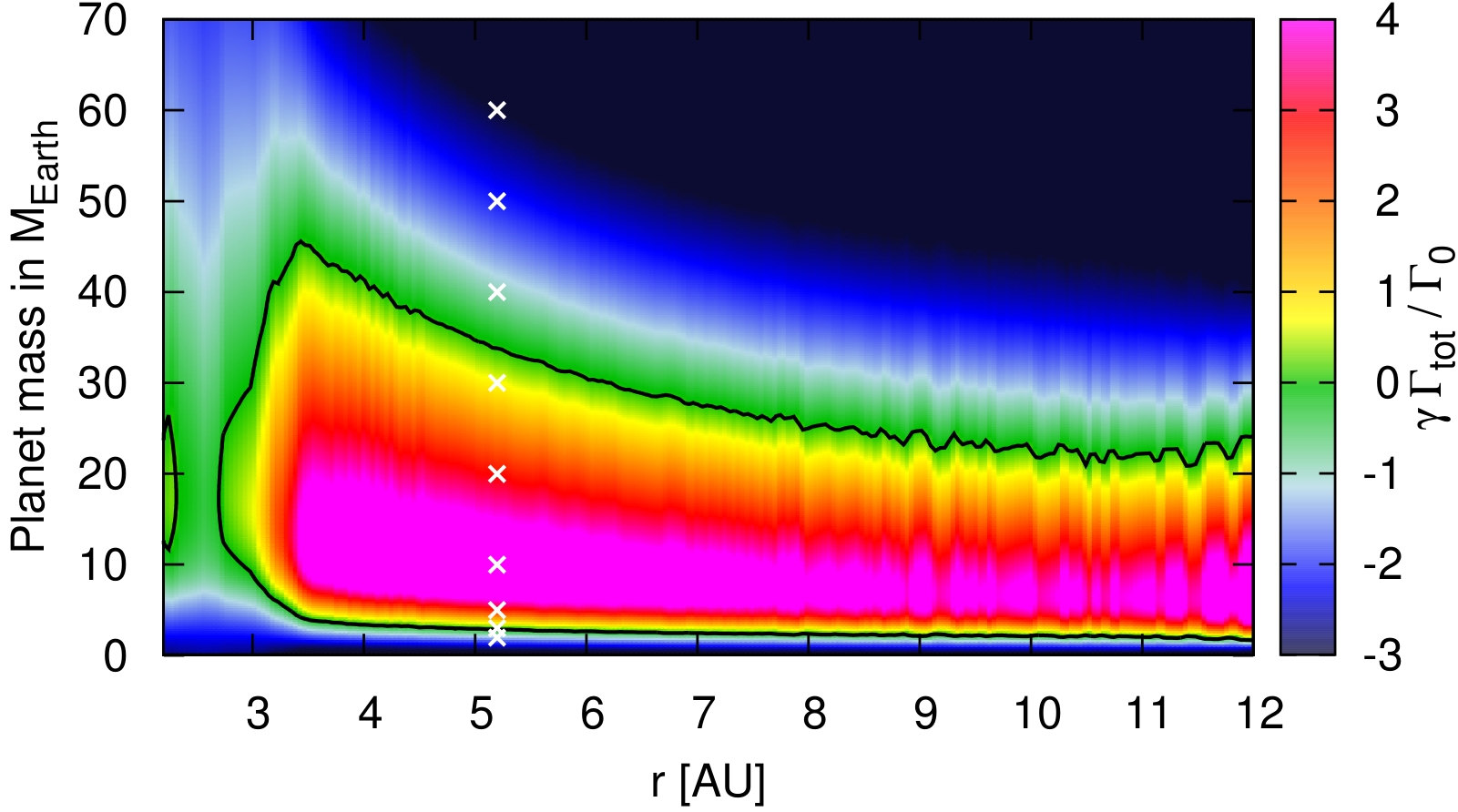}
\caption{Migration maps for simulation set  SAD with $\dot M=4\,10^{-8}\mathrm{M_\odot/yr}$ (top), SED (middle). For  comparison, the migration map of the radiative equilibrium
disc  (RED) is also plotted (bottom). The white x symbols indicate the position and the masses of the planets for which we compute the total torque with 3D simulations in this paper.}
\label{map}
\end{figure}

\section{Equilibrium discs}
In the  SED  we have embedded planets of different masses
held on  a fixed orbit at  $r=4\mathrm{AU}$,  for which we
 expect outwards migration from the torque formula
(Fig.\ref{map}, middle panel). 
Let us notice that, in the inner part of the disc,
 the aspect ratio of  the SED
is very similar to that of the RED 
(Fig.\ref{hr});  since the density gradient is the same for both discs
 the computation of the total torque  should give results similar to those 
obtained in \citet{KBK09} and \citet{LCBM14}.

\begin{table*}
\begin{center}
\begin{minipage}{100mm}
\begin{tabular}{|llll|}
\hline\noalign{\smallskip}
mass ($M_{\earth} $) & ($N_r$,$N_\theta,N_\varphi$) & $n$ cells in $x_{hs2D}$  & \\
\noalign{\smallskip}\hline\noalign{\smallskip}
\hline\noalign{\smallskip}
  $8$ & $(1326,66,908)$ & 4 & \\
  $10$ & $(1186,66,812)$ & 4 & \\
  $20$ & $(1086,82,740)$ & 5 & \\
  $25$ & $(1010,66,702)$ & 4 & \\
  $25$ & $(1238,80,876)$ & 5 & \\
  $30$ & $(911,66,620)$  & 4 & \\
  $40$ & $(770,66,542)$  & 4 & \\
 $50$ & $(670,60,468)$   & 4 & \\
 $60$ & $(612,60,430)$  & 4 & \\
 $60$ & $(764,66,540)$  & 5 & \\
 
\noalign{\smallskip}\hline
\end{tabular}
\caption{Simulations parameters for set SED. Embedded planets are at distance
$ 4\mathrm{AU}$ from the star.}
\label{table:tab1} 
\end{minipage}
\end{center}
\end{table*}

\par
When considering planets at $4\mathrm{AU}$ we observe (Fig.\ref{compaSS}) that for masses larger than $30$ Earth masses the torque is slightly larger than the one provided by the formula and the transition to negative torque occurs at $~45\mathrm{M_{\earth}}$ instead of $40\mathrm{M_{\earth}}$ as expected from the formula (Fig.\ref{map}, middle panel). However, the overall picture is quite in good agreement with the results provided by the analytical formula.  For comparison we have extend our previous study of the RED from \citet{LCBM14} for planets at $5.2 \mathrm{AU}$ towards planets with masses up to 60 Earth masses. We can appreciate in Fig.\ref{compaMNRAS} the agreement
between the results of numerical simulations and the values given by the analytical formula.
\par
In both the RED and SED, we provide  in the following subsections  a quantitative comparison between the values of the torque resulting from numerical simulations and the values given by the formula.

\subsection{Planetary masses smaller than $10M_{\earth}$}
In the case of the SED disc, the minimum planetary mass to have outwards migration is somewhat larger than that predicted in the \citet{Paaretal11} formula ( about $8 \mathrm{M_\oplus}$ against $4 \mathrm{M_\oplus}$). This minimum mass is very important for giant planet formation models. The larger transition mass that we find can be explained  with the cold finger effect studied in \citet{LCBM14}.
The unperturbed temperature at the planet location is of $112\mathrm{K}$, and, according to the criterium given in \citet{LCBM14}, a negative contribution to the total
torque (cold finger) exists  when  the Bondi radius $R_B$
is smaller that the Hill radius $R_H$. We have suggested:  
\begin{equation}
{R_B \over R_H} = {{(m_p/7.1M_{\earth})^{2/3}} \over {(a/ 5.2\mathrm{AU})(T/ 75\mathrm{K})}},
\label{coldfinger}
\end{equation}
Applying Eq.\ref{coldfinger} for planets at 4AU { where $T=112$ K}, we obtain ${R_B \over R_H} = 1$ for a $10 \mathrm{M_{\earth}}$  and ${R_B \over R_H} = 0.6$ for a $5 \mathrm{M_{\earth}}$ planet.
Thus, the expected transition between outwards and inwards migration occurs in the interval $[5,10] \mathrm{M_{\earth}}$ in very good agreement with the value of $8 M_\oplus$ found in the simulations. { In the RED disc,  the minimum planetary mass to have outwards migration is  about $4\mathrm{M_{\earth}}$ against $2\mathrm{M_{\earth}}$, due to the cold finger  effect found in \citet{LCBM14}.}

\subsection{Planetary masses in the interval $10-30 \mathrm{M_{\earth}}$}
In Fig.\ref{compaMNRAS} we notice that the maximum value of the computed torque as a function of planetary mass is  about a factor 2 smaller than the maximum provided by the formula. 
The planetary mass for which we measure the largest positive torque value is respectively of $20 \mathrm{M_{\earth}}$ for the RED case and of $30 \mathrm{M_{\earth}}$ for
the SED case. In both cases the formula provides the largest positive torque for a planet of $10 \mathrm{M_{\earth}}$, {   the onset of saturation occurs at planetary masses smaller than what found in 3D simulations. We will
discuss this point in Section 6.}
These quantitative discrepancies have already been pointed out by 
\citet{KBK09} caused by differences between 2D and 3D effects.

\subsection{Planetary masses larger than $30\mathrm{M_{\earth}}$}
The analytical formula was derived for small mass planets ($5\mathrm{M_{\earth}}$), i.e. for planets that only slightly perturb the disc.  However, larger mass planets start opening a partial gap  in the disc, meaning they significantly perturb the disc. These perturbations make the application of 
 the torque formula  dubious for masses larger than $30 \mathrm{M_{\earth}}$. However,  the density  in the corotation region is perturbed by less than $30\%$ up to $60 \mathrm{M_\oplus}$; i.e we are still far from planetary masses which really open a gap in the disc.\par 
We remark that the formula and simulations results  are in very good  agreement for masses larger than $30 \mathrm{M_{\earth}}$.  The torques in the  formula are calculated from the gradients of surface density, temperature and entropy in the unperturbed disc.
In the formula, the corotation torque saturates as the planet mass increases, making the total torque transit to negative. In our simulations, the total torque is obtained directly from the density structure of the disc, meaning that on top of saturation effects, the partial opening of a gap is also taken into account, decreasing even more the corotation torque.
 That these two different approaches match in the total torque may be a coincidence, but we remind that the density in the corotation region is perturbed by less than $30\%$ up to $60 \mathrm{M_\oplus}$.

\begin{figure}
\includegraphics[height=6.5truecm,width=6.5truecm]{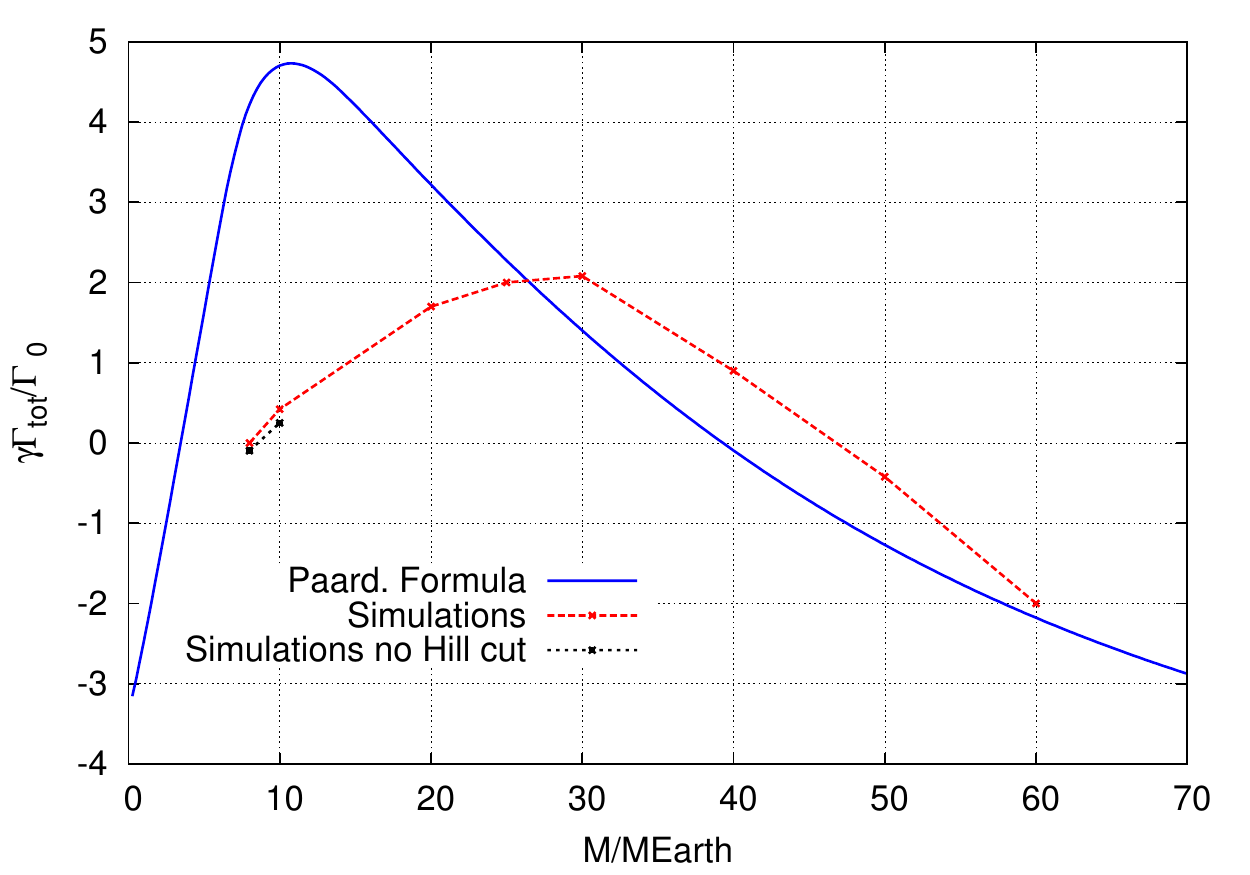}
\caption{Stellar irradiated equilibrium disc (SED): comparison with \citet{Paaretal11} formula for  planets on fixed orbits  at $4\mathrm{AU}$. For planets of $10$ and $8\mathrm{M_{\earth}}$  we also show the value of the torque computed without the Hill cut (see text).}
\label{compaSS}
\end{figure}

\begin{figure}
\includegraphics[height=6.5truecm,width=6.5truecm]{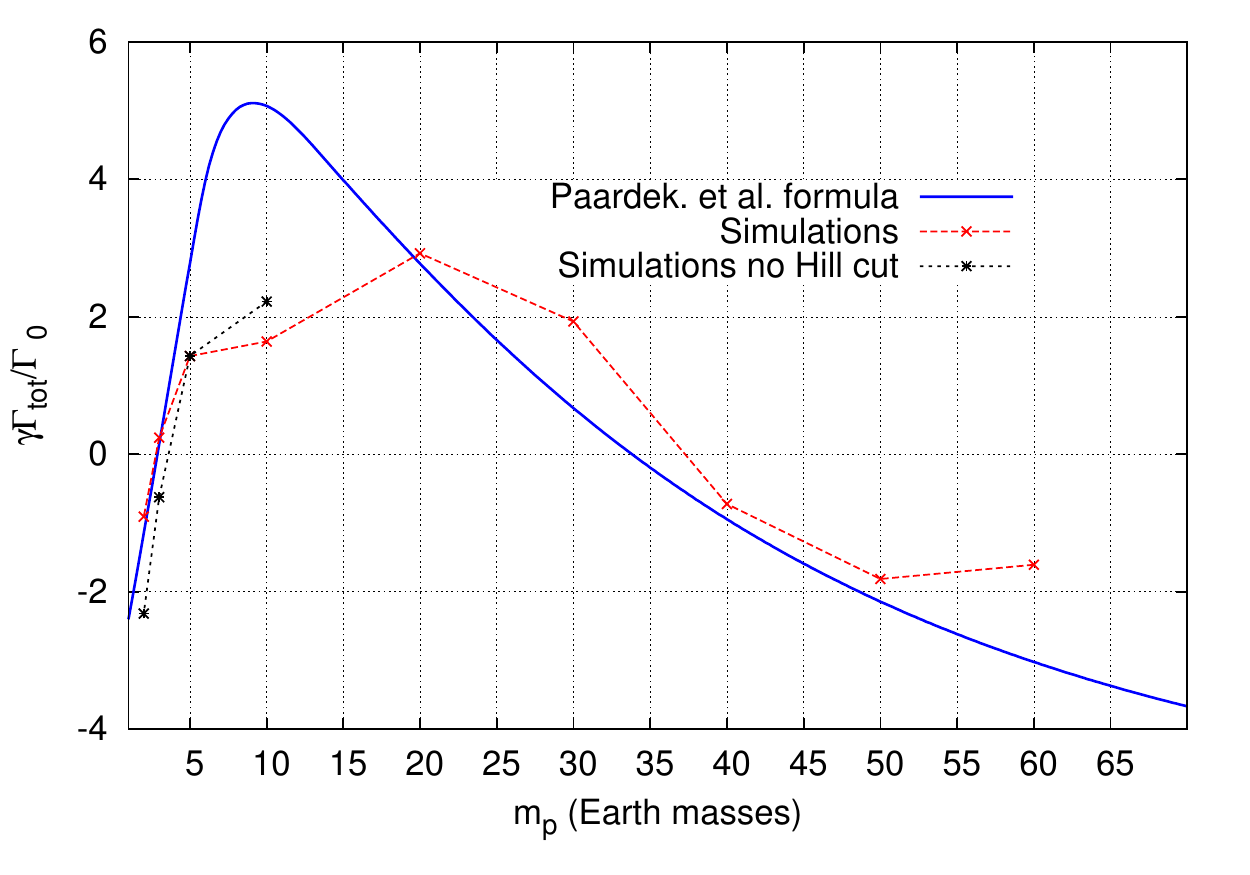}
\caption{Same as Fig.3 for the  radiative disc (RED) studied in \citet{LCBM14} whose aspect ratio is reported in Fig.\ref{hr} and the corresponding migration map in Fig.\ref{map}, bottom panel. For  planet of mass smaller than $10 \mathrm{M_{\earth}}$  we also show the value of the torque computed without the Hill cut (see text).}
\label{compaMNRAS}
\end{figure}

We can conclude this section saying that when we take into account  stellar irradiation in an equilibrium disc no additional effects on the total torque acting on planets in the explored mass range of $[8,60]$ Earth masses are observed.
 Moreover, the analytical formula \citet{Paaretal11} can be used with good confidence when performing, for example, N-body simulations with a migration prescription.

\section{Stellar irradiated accretion discs}

Protoplanetary discs accrete gas onto the central Star \citep{LP74}. Thus equilibrium discs like the SED and RED cases investigated above are not a good approximation of reality, given that they do not transport mass radially. A much better approximation are the discs where the mass flow $\dot{M}$ is independent of radius, called accretion discs. Their structure has been investigated in \citet{PaperII}. Here we check the validity of the \citet{Paaretal11} formula in these more realistic discs. 

\par
We consider the case of an accretion disc with constant $\dot M=4\, 10^{-8}\mathrm{M_\odot/yr}$ and we  compute the total torque for planets in the mass interval $20-70$ Earth masses placed at distances of $4\mathrm{AU}$, $5\mathrm{AU}$ and $6\mathrm{AU}$ from the central star. We have also made a few tests on smaller planetary masses. The values of the masses and resolutions $(N_r,N_\theta,N_\varphi)$ for the SAD disc are shown in Table \ref{table:tab2}. 

\begin{table*}
\begin{center}
\begin{minipage}{100mm}
\begin{tabular}{|lllll|}
\hline\noalign{\smallskip}
mass ($M_{\earth} $) & ($N_r$,$N_\theta,N_\varphi$) & $n$ cells in $x_{hs2D}$  & distance (AU)  \\
\noalign{\smallskip}\hline\noalign{\smallskip}
\hline\noalign{\smallskip}
 $5$ & $(1600,110,1132)$ & 4 & 4 & \\
   $15$ & $(1320,96,910)$ & 4 & 4& \\
   $15$ & $(1256,90,886)$ & 6 & 5 & \\
  $20$ & $(1066,74,750)$ & 4 & 4 & \\
  $20$ & $(1600,100,1132)$ & 7 & 4 & \\
  $20$ & $(1318,80,920)$ & 6 & 5 & \\
  $20$ & $(1528,90,1072)$ & 7 & 5 & \\
  $20$ & $(1134,90,758)$ & 7 & 6 & \\
  $30$ & $(1502,90,1068)$ & 7 & 4 & \\
  $30$ & $(1250,76,878)$ & 7 & 5 & \\
   $30$ & $(1080,78,718)$ & 7 & 6 & \\
  $40$ & $(1360,86,972)$ & 7 & 4 & \\
  $40$ & $(1080,70,758)$ & 7 & 5 & \\
   $40$ & $(926,66,622)$ & 7 & 6 & \\
 $50$ & $(1226,76,868)$ & 7 & 4 & \\
  $50$ & $(966,70,680)$ & 7 & 5 & \\
   $50$ & $(826,66,558)$ & 7 & 6 & \\
  $60$ & $(1118,80,794)$ & 7 & 4 & \\
  $60$ & $(882,66,624)$ & 7 & 5 & \\
   $60$ & $(754,66,514)$ & 7 & 6 & \\
  $70$ & $(1034,74,730)$ & 7 & 4 & \\
  $70$ & $(808,66,576)$ & 7 & 5 & \\
   $70$ & $(692,56,476)$ & 7 & 6 & \\
 
\noalign{\smallskip}\hline
\end{tabular}
\caption{Simulations parameters for set SAD.}
\label{table:tab2} 
\end{minipage}
\end{center}
\end{table*}
 \par

 We have used a resolution of 7 grid-cells in the half width of the horseshoe region for all the
computations in the interval $20-70$ Earth masses.
We have checked that this  is needed in order to the have results independent of the resolution.
 For small planetary masses  a resolution of 4 grid-cells is enough to achieve convergence (see table \ref{table:tab2}).

\par
In Fig.\ref{compaMd} (top) we report the total torque computed on planets held 
on fixed orbits at $4\mathrm{AU}$.  The transition from inwards to outwards migration 
 of small masses occurs for planet masses about twice as big as predicted by the Paardekooper et al. formula.   We find again the ``cold finger'' effects, with a transition to negative torque at about $10$ Earth masses. 

\par
In the interval $[15,30]\mathrm{M_{\earth}} $ we observe a positive total torque with values smaller than expected from the formula as in the previously examined RED and SED. In the interval $[40,70]\mathrm{M_{\earth}} $ the results from numerical simulations very nicely agree with the torque provided by the analytical formula. We  find positive torques up to planetary masses of $60\mathrm{M_{\earth}}$.  

\par
Similar results are found at $5\mathrm{AU}$ and $6\mathrm{AU}$ (Fig.\ref{compaMd}, middle and bottom panels),
{ precisely, cores of $50 M_{\earth}$ undergo outwards migration at  $5\mathrm{AU}$ and at $6\mathrm{AU}$. From the migration map 
the transition between outwards and inwards migration occurs for 
slightly larger masses; however  the overall picture and the radial extent
of the outwards migration is quite well  reproduced.}
\par { As we pointed out in Section 3} this is an important result for models of giant planet formation. This may solve the problem pointed out in \citet{ColeNels14}   where all planets were lost by migration before becoming gas-giant planets. However, we stress that this is true only for discs with large $\dot{M}$ like the one studied here. As pointed out in \citet{PaperII,BJLM15}, the maximum planet mass for outwards migration decreases with decreasing $\dot{M}$.

 \begin{figure}
\includegraphics[height=6.5truecm,width=7truecm]{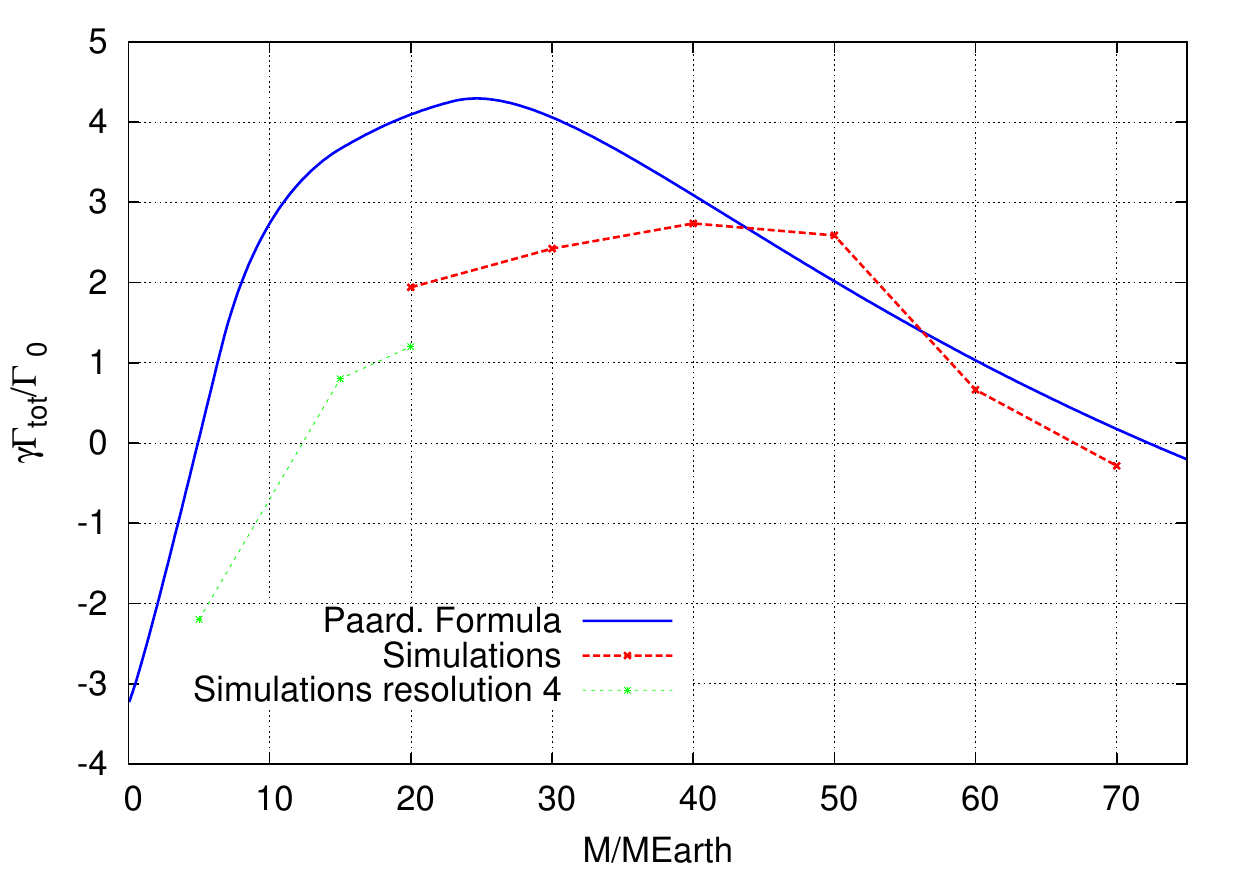}
\includegraphics[height=6.5truecm,width=7truecm]{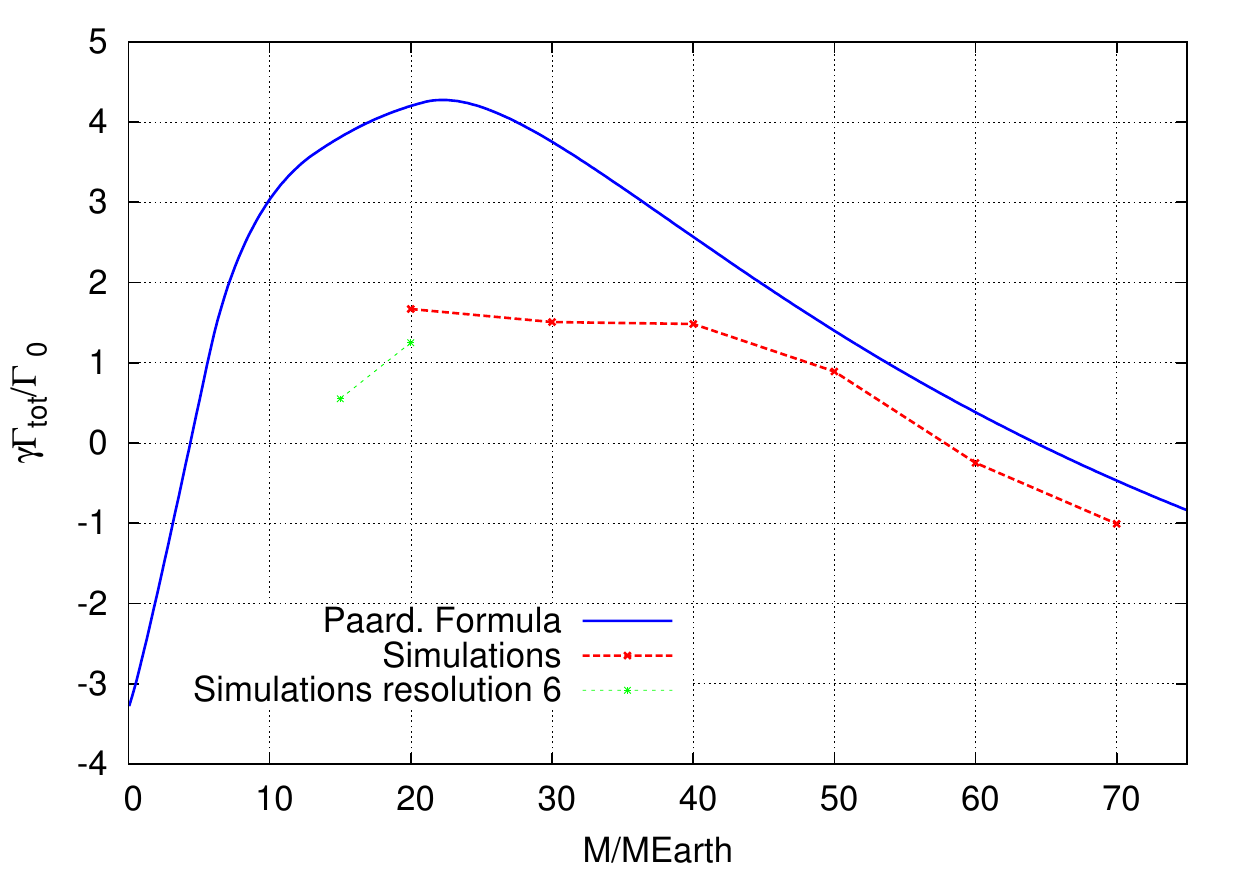}
\includegraphics[height=6.5truecm,width=7truecm]{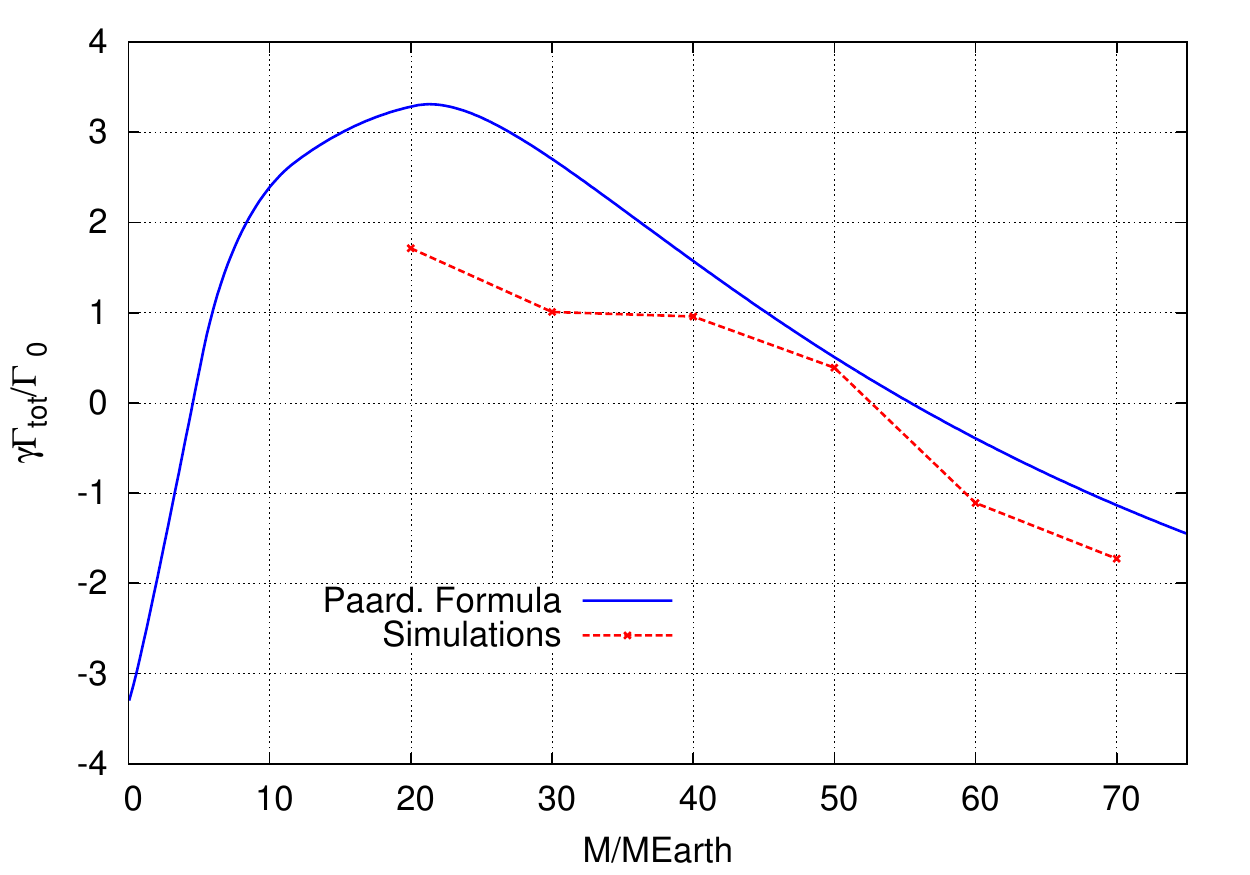}
\caption{Accretion disc: comparison with Paardekooper et al. (2011) formula. {\bf Top panel}: planets at $4\mathrm{AU}$, {\bf middle}: planets at $5\mathrm{AU}$ {\bf bottom panel}: planets at $6\mathrm{AU}$. The red points are obtained with a resolution of 7 grid cells in the half width of the horseshoe region, the green ones with a resolution of 4 grid cells. }
\label{compaMd}
\end{figure}

\subsection{Radial torque distribution}

To further confirm that the outwards migration detected in the accretion disc
corresponds to that previously found in radiative discs without stellar heating  and without a net mass flux through the disc we show  the contribution to the  torque in the close vicinity of the planets for both SAD and RED.
We compute the radial torque density
$\Gamma (r)$, i.e  the torque exerted on the planet by a ring of disk material located at a distance $r$ from the star. 
The integral of $\Gamma(r)$ on the radial coordinate provides the total
torque $\Gamma_{tot}$ shown in the previous section:
\begin{equation}
\Gamma_{tot}=\int _{r_{min}}^{r_{max}} \Gamma(r)dr \ .
\end{equation}
 Fig.\ref{normtorque}  shows the radial distribution of the torque exerted on  planets of different masses in the accretion disc with $\dot M=4\,10^{-8}M_\odot/yr$ at $5\mathrm{AU}$ (top panel) and the radial torque density for the RED disc for planets placed at $5.2 \mathrm{AU}$ (bottom panel).
 
\par
In a disc without thermal effects (isothermal disc) $\Gamma(r)$ would be positive for $r<r_p$ and negative for $r>r_p$. Moreover, $\Gamma(r)/ \Gamma_0$  would be independent of the planet's mass. Here we observe that $\Gamma(r)/\Gamma_0$ changes with the planetary mass. This effect is due to the saturation of the entropy-driven corotation torque  \citep{BarMass08,KBK09} which is a function of planetary masses.  We remark in Fig.\ref{normtorque} that, when the planetary mass increases from $20$ to $70 M_\oplus$, the absolute value of $\Gamma/\Gamma_0$ decreases and the location where $\Gamma$ transits from positive to negative, which is at $r>r_p$ for the $20 M_\oplus$ planet, approaches $r_p$. The displacement of the point where $\Gamma(r)=0$ relative to $r_p$ is indicative of the strength of the total positive torque. In this case, the total torque decreases for increasing planetary mass, until it becomes slightly negative for the $70 M_\oplus$ planet.  
In the case of the RED disc the results are very similar, although one can notice quantitative differences. In this case, $\Gamma(r_p)=0$ for a planet of
 $40 M_\oplus$. In fact, the transition to inwards migration occurs  at about $40 M\oplus$, as shown in Fig. \ref{compaMNRAS}.

\begin{figure}
\includegraphics[height=6.5truecm,width=6.5truecm]{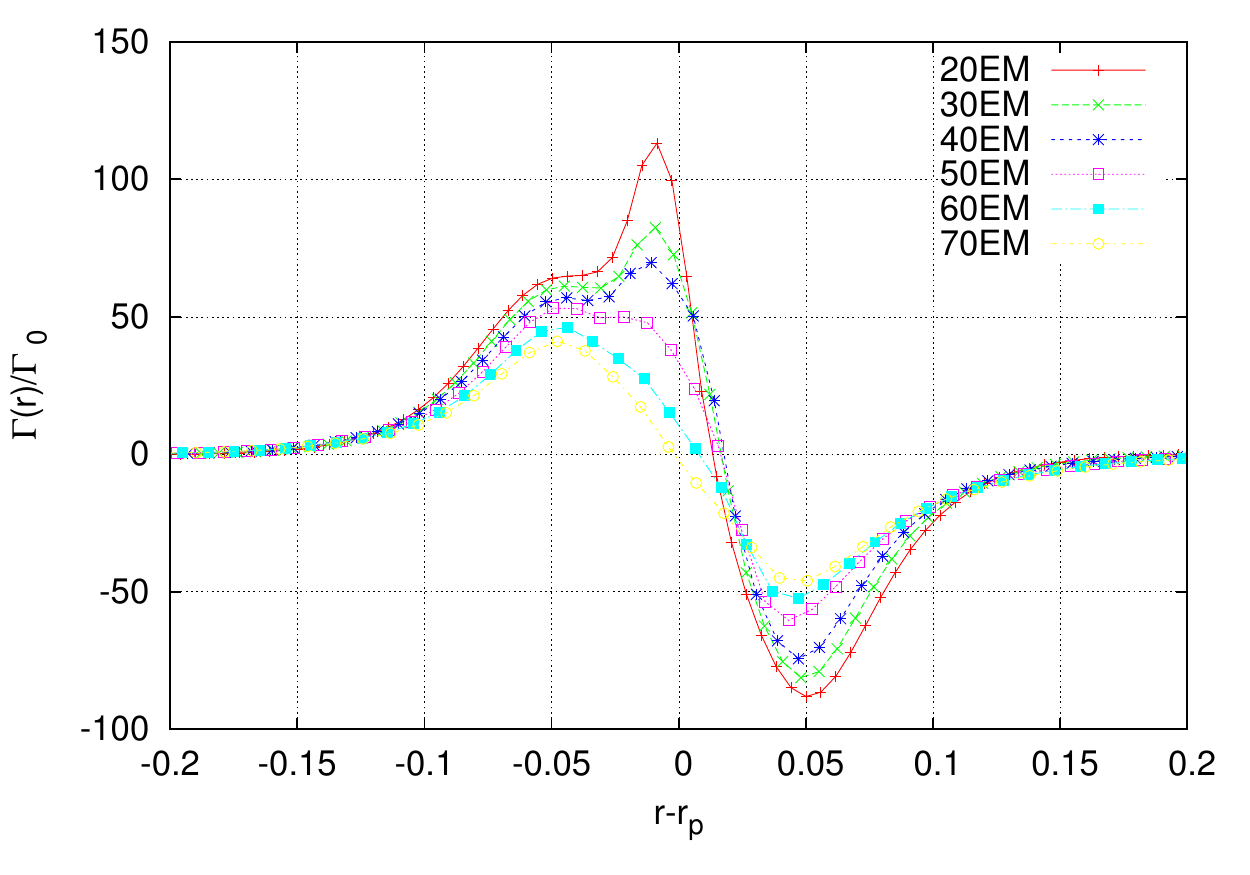}
\includegraphics[height=6.5truecm,width=6.5truecm]{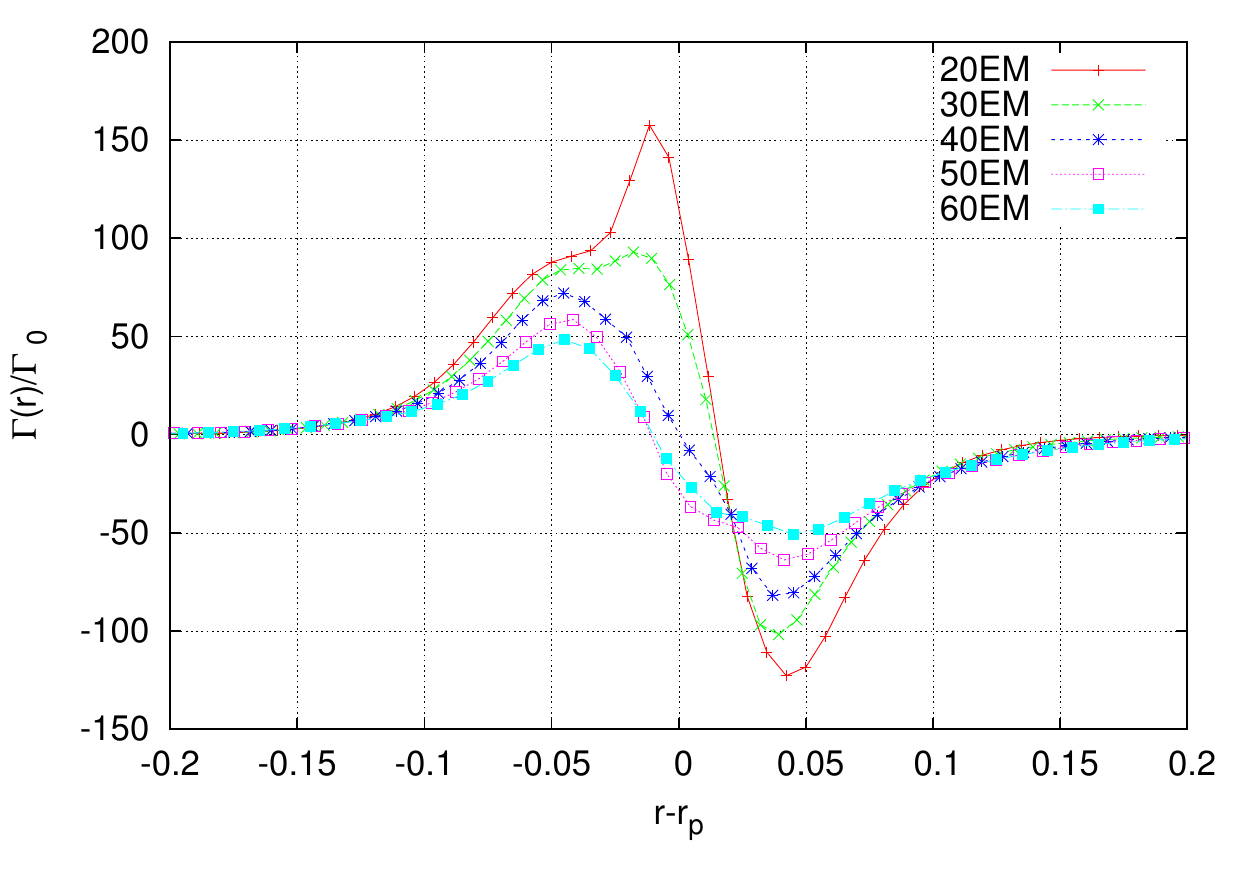}
\caption{Radial torque density for planet of different masses at $5\mathrm{AU}$ in the accretion disc with $\dot M=4\,10^{-8}M_\odot/yr$ (top) and in the RED disc for planets at $5.2\mathrm{AU}$ (bottom). }
\label{normtorque}
\end{figure}

\section{On the width of the horseshoe region}
We now go beyond the raw comparison of the torques predicted in Paardekooper et al. formula and measured numerically, in order to understand how such a good agreement is achieved for large masses but not for low masses. This is indeed surprising because the formula is derived in principle in the linear regime, while for $m_p>30M_{\earth}$, where the agreement is best, the non-linearities should start to appear (because of the condition $q>h^3$). Here we focus on the size of the corotation zone, which is a key parameter for the estimate of the corotation torque and of the torque saturation. 
\par
According to \cite{MassetDangelo06}  Eq.\ref{xhs}, obtained for 2D discs, is valid only for planet to star mass ratio $q< h^3$. The horseshoe region for large masses ($q>1.5\, 10^{-4}$ in \cite{MassetDangelo06}) behaves as in the restricted three body
 problem (RTBP, hereafter), i.e. scales with $q^{1/3}$.  For intermediate masses, the width of the horseshoe region is
larger than what predicted by the  $q^{1/2}$ scaling. This is a manifestation of the flow nonlinearity turning out in a boost of the corotation torque (\cite{MassetDangelo06}). \par
 In this section, we provide a measure of the half width of the horseshoe region, ( $x_{hs3D}$ hereafter), for SED and for the SAD with planets respectively at $4\mathrm{AU}$ and  at $5\mathrm{AU}$ and compare it to $x_{hs2D}$ of Eq.\ref{xhs}. 
Since differences with respect to Eq.\ref{xhs} can come from
3D effects as well as from radiative effects 
 we will proceed in 2 steps: i) we determine the difference between the horseshoe width in 2D and 3D simulations in an isothermal setting, ii) we compare isothermal 3D to fully radiative 3D. 
\subsection{Comparison between 2D and 3D simulations}
For the two dimensional case we used the FARGO code (\cite{Masset00}). We recall that in 2D a softening length $\epsilon$ is  applied to the planet
 potential through:
\begin{equation}
\Phi = -{Gm_p \over \sqrt{a_p^2+\epsilon^2}}
\label{pot2D}
\end{equation}
In 2 dimensional models, Eq.\ref{pot2D} allows to mimic the average influence  that the planet
would have  on the vertical gas column.
The measure of the width of the horseshoe region is determined by computing the streamlines.
In Fig.\ref{xhs2D} we plot the half width of the horseshoe region normalized over the planet semi-major axis $a_p$ for different values of the planet to star mass ratio $q$, in the mass range considered in this paper ($[10,70]M_{\earth}$).  
Using  the nominal setting given in \cite{MassetDangelo06} we recover their results (see their Fig.9).
In Fig.\ref{xhs2D}  the half width of the horseshoe region obtained with
a softening length of $\epsilon=0.3H$ (as in  \cite{MassetDangelo06}) is fitted by
 \begin{equation}
xM_{hs2D}\simeq 2.45a(q/3)^{1/3}
\label{xhsM}
\end{equation}
 for $q>1.35\,10^{-4}$ 
while the  $x_{hs2D}$ law of Eq.\ref{xhs} nicely fits the data for $q\leq 4.5\times 10^{-5}$. The value of the aspectratio for these simulations is $h=0.044$,
this value is used in the plot of the $x_{hs2D}$ law.\par
We remark that in 2D  the effect of the softening is felt well outside the Hill radius, so that the horseshoe region clearly depends  on the softening length.
In all our 3D numerical simulations we have used a softening length
of $0.6R_{H}$.
  In order to compare 2D and 3D isothermal results we
 have also run a set of 2D simulations with a softening length of 
$\epsilon=0.6H$. In Fig.\ref{xhs2D}  we observe that in this case the transition between the $q^{1/2}$ and the  $q^{1/3}$ scaling occurs in the mass interval $9\,10^{-5}<q<1.65\,10^{-4}$.
The width of the horseshoe region, both in the linear and in the RTBP regime, is reduced of about  $10\%$ with respect to the values obtained for the smaller softening length. 
In Fig.\ref{xhs2D} we have also reported the measure of the width of the horseshoe region  obtained with 3D isothermal simulations  with $\epsilon = 0.6R_{H}$.
The  disc parameters are the ones of SAD set, with planets at $5\mathrm{AU}$. 
The aspectratio at this distance from the star is $h=0.044$ (Fig.\ref{hr}). 
For 3D simulations the measure  of the width of the horseshoe region is determined by computing the streamlines on the disc midplane.  Results are in good agreement with the 2D case with $\epsilon=0.6H$  \footnote{A value of the softening length of $\epsilon=0.7H$,  is shown to provide a radial  torque density  very similar to the one obtained with 3D simulations (see \cite{Muller12}.)} so that we do not observe 2D versus 3D effects in the horseshoe region width.\par

\begin{figure}
\includegraphics[height=6.5truecm,width=7.5truecm]{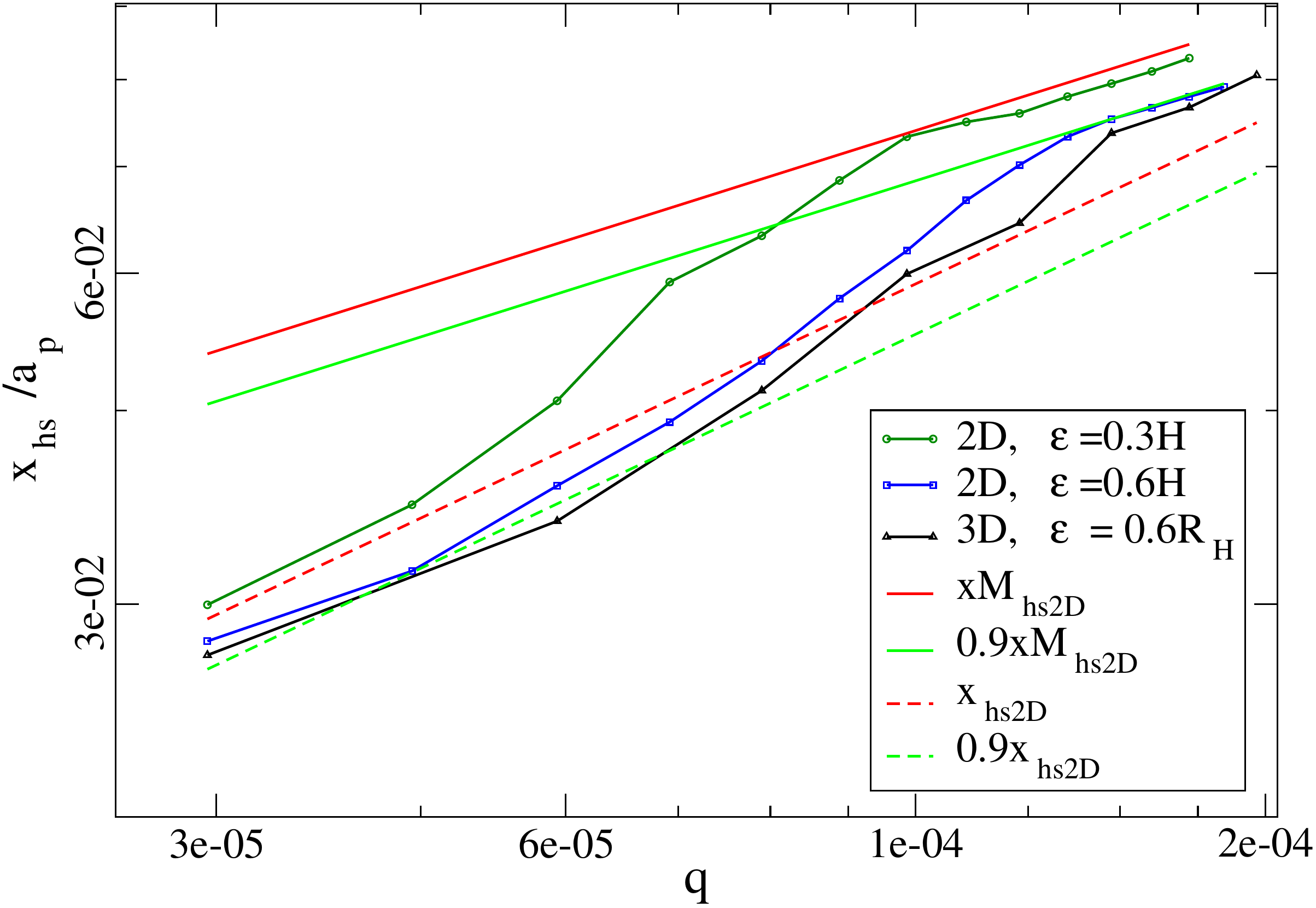}
\caption{Simulations with isothermal EOS. Half-width of the horseshoe region normalized over $a_p$ as a function of the planet to star mass ratio $q$. Results are shown for two sets of 2D simulations with respectively $\epsilon=0.3H$ and $\epsilon=0.6H$ and for 3D simulation with isothermal equation of state and $\epsilon = 0.6R_{H}$.  The lines show the different fits in $q^{1/2}$ ( with labels $x_{hs2D}$ and $0.9x_{hs2D}$, see text) and  $q^{1/3}$ (with labels $xM_{hs2D}$ and $0.9xM_{hs2D}$, see text).}
\label{xhs2D}
\end{figure}

\subsection{Comparison between 3D isothermal and 3D radiative simulations}
We call $xM_{hs3D} \simeq 0.9xM_{hs2D}$ the width of the horseshoe
region obtained by the fit of 3D isothermal simulations in the RTBP regime
and   $x_{hs3D} \simeq 0.9x_{hs2D}$ the  fit obtained for the linear regime.
Fig.\ref{xhs3D} shows the measure of the width of horseshoe region
for SED with planets at 4AU \footnote{The aspectratio  for SED at 4AU is $\tilde h=0.041$ (Fig.\ref{hr}). To be compared to the other data sets having $h=0.044$ a
 rescaling by $\sqrt{ \tilde h/ h}$ is applied.}, and SAD with planets at 5AU. 

For comparison we plot the results of the 3D isothermal simulations of 
Fig.\ref{xhs2D}.
We recover a regime with a $q^{1/2}$ scaling  and a  regime with a $q^{1/3}$ scaling and in both regimes the horseshoe region is narrower than in the isothermal case of about a factor $\gamma^{-1/4}$ as expected from \citet{PaaPap09}.
In the same Fig.\ref{xhs3D} we have plotted the law for the horseshoe region used in the Paardekooper formula, i.e. $x_{hs2D}\gamma^{-1/4}$.

\begin{figure}
\includegraphics[height=6.5truecm,width=7.5truecm]{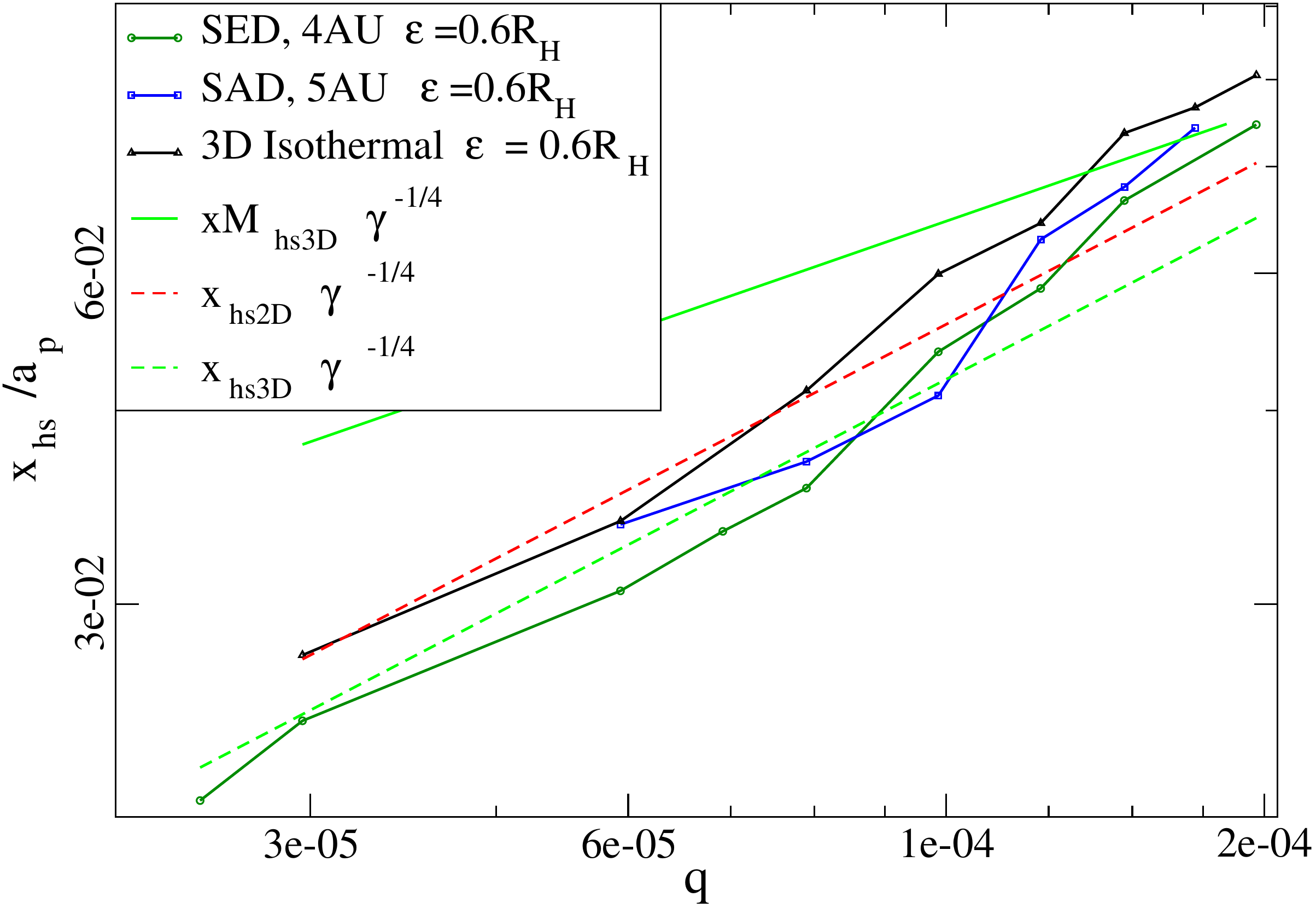}
\caption{Same as Fig.\ref{xhs2D} for 3D simulations for respectively SED with planets at 4AU,
and SAD with planets at 5AU and for the 3D isothermal run of Fig.\ref{xhs2D}.
 The lines show the  fits in $q^{1/2}$ and  $q^{1/3}$. The law for the horseshoe region used in the Paardekooper formula, $x_{hs2D}\gamma^{-1/4}$, is also shown.}
\label{xhs3D}
\end{figure}
}

We observe that the  measured horseshoe width is respectively $\simeq 10\%$ 
narrower than  $x_{hs2D}\gamma^{-1/4}$ in the linear (or $q^{1/2}$ scaling) regime
and about $\simeq 10\%$ larger in the nonlinear (or $q^{1/3}$ scaling) regime.
Now a larger horseshoe region causes saturation earlier. We discuss this point in Section 6.4.
\par
In the isothermal case  the transition between the two regimes
is associated to a boost in the corotation torque \citep{MassetDangelo06}
and  a small softening length is required to detect it.
Therefore, we can wonder if we have missed some effects
 in the torque computation by using a softening length of $0.6R_{H}$.

\subsection{Comparison between 3D radiative discs with different softening length}
\begin{figure}
\includegraphics[height=6.5truecm,width=7.5truecm]{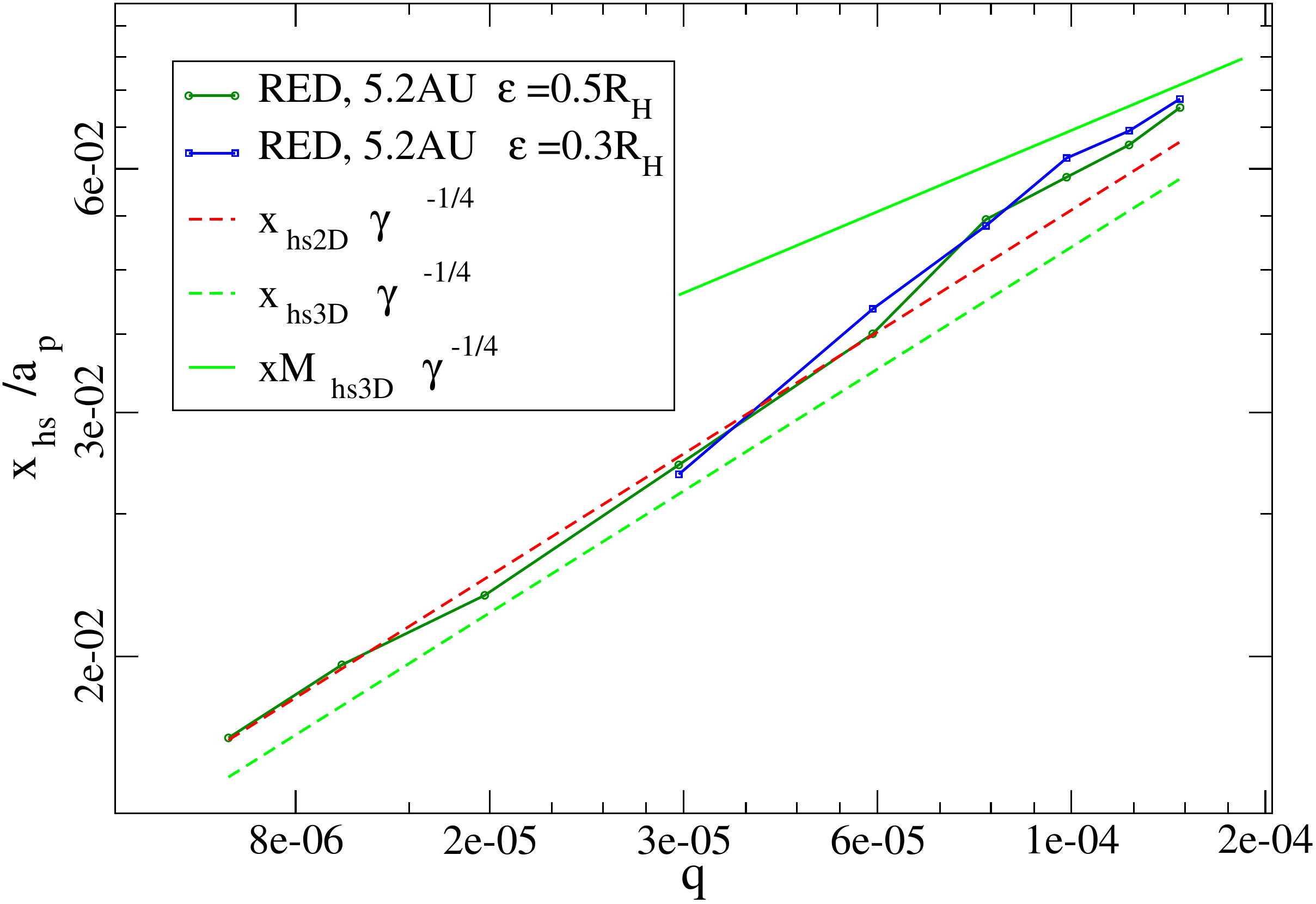}
\caption{Half-width of the horseshoe region as a function  planet to star mass
ratio ($q$) for 3D simulations of RED set, for respectively $\epsilon=0.5R_{H}$ and $\epsilon=0.3R_{H}$.
 The lines show the  fits in $q^{1/2}$ and  $q^{1/3}$ obtained for the stellar irradiated discs (Fig.\ref{xhs3D}). The law for the horseshoe region used in the Paardekooper formula, $x_{hs2D}\gamma^{-1/4}$, is also shown.}
\label{xsRED}
\end{figure}

In the case of 3D  isothermal simulations  simulations it has been shown
\citep{KBK09} that a decrease in the softening length for a planet 
of $20 M_{\earth}$ corresponds to a drastic change in the density structure in the planet vicinity which has an important impact on the corotation torque.
Precisely, the authors found a torque excess in agreement with \citet{MassetDangelo06}.
The situation changes when taking into account thermal effects. In this
case a deeper potential corresponds to an increase of temperature near the planet so that the density distribution  close to the planet is not  drastically changed. In \citet{KBK09} (their Fig.14, top ) changing the smoothing length
 from $\epsilon=0.8R_H$ and $\epsilon=0.5R_H$ in a fully radiative setting acts in increasing the torque of about $5\%$, with respect to $30\%$ of increase  in the corresponding isothermal case. 
We have extended the discussion of \citep{KBK09} concerning the dependence of the torque on the softening length by computing the width of the horseshoe region and the total torque acting on fixed planets 
 in the range $[10:60]M_{\earth}$ for the RED set  with $\epsilon = 0.3R_{H}$.
In Fig.\ref{xsRED} it appears clearly that the width of the horseshoe region is
almost not affected by the change in the softening length \footnote{The aspectratio  for RED at 5.2AU is $\tilde h=0.04$ (Fig.\ref{hr}). To be compared to 
data having $h=0.044$ a
 rescaling by $\sqrt{ \tilde h/ h}$ is applied.}. The measured torque 
is also almost not   affected by the change in the softening length. \par

 We observe that in the $q^{1/2}$ regime the horseshoe region is nicely fitted by the law  used in the Paardekooper formula (Fig.\ref{xsRED}).
However, although the measured width of the horseshoe region perfectly matches the law used in the formula
up to $20M_{\earth}$ the measured torque fits only qualitatively the one provided by the formula as explained in Section 4.2.
This is not surprising since many other parameters enter in the formula and can possibly be different in realistic 3D discs.
 \par
In isothermal discs a torque excess is correlated to the departure from the linear regime  \citep{MassetDangelo06} while 
in all our radiative discs we do not observe the same phenomenon.
Taking into account thermal effects, the transition from the linear to the nonlinear regime, even considering a
small softening length (Fig.\ref{xsRED}), is smoother in
radiative discs (Fig.\ref{xhs3D} and Fig.\ref{xsRED})
with respect the isothermal case with small softening length  (Fig.\ref{xhs2D}).
Moreover,  the mass at which we observe a  departure from the linear regime in Fig.\ref{xhs3D} and Fig.\ref{xsRED}  corresponds to the onset of torque saturation in Fig.\ref{compaSS},\ref{compaMNRAS} and \ref{compaMd}, middle panel.

\par

\subsection{Impact of the measured horseshoe width on planet migration}
The width of the horseshoe region is of crucial importance for the saturation of the torque acting on more massive planets. A larger width of the horseshoe region increases the viscous timescale (Eq. \ref{taunu}) making torque saturation easier. As demonstrated in Fig.\ref{xhs3D} and Fig.\ref{xsRED}, the width of the horseshoe region increases for more massive planets and does not follow the simple law of $x_{hs2D} \gamma^{-1/4}$, which is only valid for low mass planets.

\begin{figure}
\includegraphics[height=6.5truecm,width=7truecm]{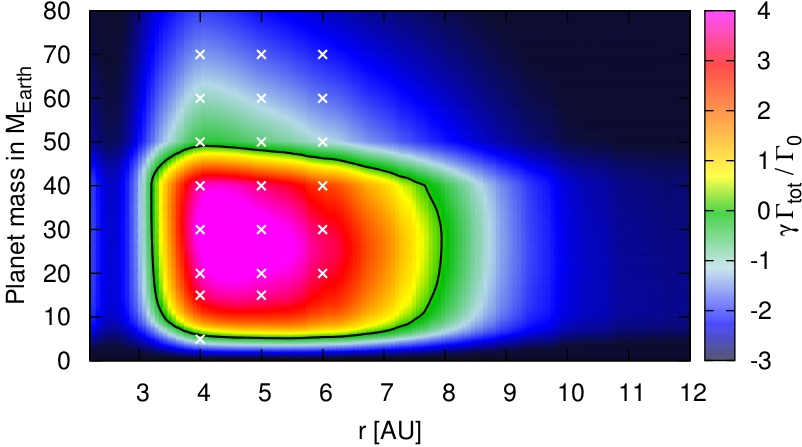}
\caption{Migration maps for simulation set  SAD with $\dot M=4\,10^{-8}\mathrm{M_\odot/yr}$ obtained using the width of the horseshoe region measured by our simulations (Fig.\ref{xhs3D}) instead of $x_{hs2D} \gamma^{-1/4}$.}
\label{maplarge}
\end{figure}

In Fig.\ref{maplarge} we show the migration map of the SAD disc, where we used the width of the horseshoe region measured by our simulations (Fig.\ref{xhs3D}) instead of $x_{hs2D} \gamma^{-1/4}$.  When comparing to Fig.\ref{map},top panel we see that
for planetary masses of up to $40 M_{\earth}$ the migration map remains unchanged while for larger planetary masses, we observe that the transition of the torque from positive to negative occurs in the interval $40-50 M_{\earth}$.
 The  transition occurs in the interval  $60-70 M_{\earth}$  in Fig.\ref{map},top panel. This is caused by a larger width of the horseshoe region, which makes saturation of the corotation torque easier. Hence the total torque transitions into negative values at smaller planetary masses. Adapting the width of the horseshoe region to our measured values in the torque formula therefore over predicts the transition to inward migration compared to our simulations.

The difference between our simulations and the torque predicted by the  formula is therefore not only caused by a different width of the horseshoe region. The cooling process plays also an important role and  the vertical cooling in 2D disc is treated like blackbody radiation, while our 3D disc features vertical heat diffusion.

Therefore we think that the quantitative agreement between the torque formula of \citet{Paaretal11} and simulations for planetary masses $m_P >30M_{\earth}$ is a coincidence and the formula should be updated with results of 3D simulations. However, this is beyond the scope of this paper and left for future work.

\section{Conclusion}
Using 3D hydrodynamical simulations including stellar and viscous heating as well as radiative cooling we have computed the torque acting on 
planets of various masses kept on fixed  circular orbits in both equilibrium and accretional protoplanetary disks.  We confirm the results previously obtained 
on constant viscosity equilibrium discs, where the only source of heating was  provided by viscous friction. The comparison between the total torque obtained with numerical simulations and the one predicted by semi-empirical formula  \citep{Paaretal11} is in quantitatively good agreement for large masses ($>20\mathrm{M_{\earth}}$). 
In particular, the formula predicts well the maximum planet mass at which 
inwards migration is prevented by the action of the entropy-driven corotation torque.

Instead, although our results remain similar to those expected by  the Paardekooper formula, we observe quantitative differences as large as a factor of $\sim 2$ concerning: (i) the value of the maximum positive torque as a function of {the} planet's mass, (ii) the value of the planet mass at which the torque is  maximal and (iii) the minimum planet mass allowing outwards migration. The last difference is due to the presence of a "cold finger effect" already discussed extensively in \citet{LCBM14}.
The  differences concerning the items (i) and (ii) 
have already been pointed out by 
\citet{KBK09} as caused by differences between 2D and 3D effects. 
{ Moreover, by measuring the width of the horseshoe region we have observed that
the   value of the planet mass at which the torque is  maximal is correlated to the departure from the linear regime.}
For large planetary masses the  quantitatively good agreement between the numerical results and the prediction of the analytic formula was unexpected since the formula is based on unperturbed discs while planets with masses larger than $20\mathrm{M_{\earth}}$ start opening a partial gap.  { Precisely, torque saturation has the same behavior in the torque formula and in simulations though saturations effects appear for larger planetary masses with respect to the torque formula. From the measure of the width of the horseshoe region, we have found that the onset of saturation  is correlated to the onset of the nonlinear regime not taken into account by the formula.
 Plugging the measured horseshoe width in the formula
 gives a worse comparison. Therefore, we think that the quantitative agreement
on large masses is a coincidence, and future works is needed
to provide a more accurate torque expression.}
\par
{ At the present state of the art,} though there are some differences between the torque formula and 3D simulations, the zero torque location at the transition  from outwards to inwards migration at large planetary masses is reproduced 
surprisingly well, making the formula applicable to the study of planetary migration properties from the unperturbed disc structure.

\section*{Acknowledgments}
{ We thank an anonymous referee for raising  important points on the relation between the width of horseshoe region and the corotation torque.}
The Nice group is thankful to ANR for supporting the MOJO project
(ANR-13-BS05-0003-01).
This work was performed using HPC resources from GENCI [IDRIS]
(Grant 2014, [100379]). { B. Bitsch thanks the Knut and Alice Wallenberg Foundation for their financial support.}

\end{document}